\documentclass{XVIII_RiNEm_Proc}
 \def\beq{\begin{equation}}
 \def\eeq{\end{equation}}
 \def\beqa{\begin{eqnarray}}
 \def\eeqa{\end{eqnarray}}
 \def\beqa*{\begin{eqnarray*}}
 \def\eeqa*{\end{eqnarray*}}
 \def\barr{\begin{array}}
 \def\earr{\end{array}}

 \def\btabular{\begin{tabular}}
 \def\etabular{\end{tabular}}
 \def\btable{\begin{table}}
 \def\etable{\end{table}}
 \def\<{\langle}
 \def\>{\rangle}

\title{Open Cell Metal Foams for Beam Liners ?}
\author{R.P. Croce, S. Petracca, A. Stabile\supscr{*}}
\affiliation{
WavesGroup, University of Sannio at Benevento, 82100 Benevento, IT
}

\mailaddr{
\href{mailto:stabile@sa.infn.it}
{\supscr{*}arturo.stabile@sa.infn.it}
}

\begin{document}
\maketitle
\begin{abstract}
The possible use of open-cell metal foams for particle accelerator beam liners
is considered. Available materials and modeling tools are reviewed,  
potential pros and cons are pointed out, and a study program is outlined.
\end{abstract}

\section{INTRODUCTION}
\label{sec:intro}
Molecular gas desorption from the beam-pipe wall 
due to synchrotron radiation
should be properly taken into account 
in the design of high energy particle accelerators and storage rings.
This is specially true for hadron colliders, where nuclear scattering in the
residual gas, besides limiting the beam luminositity lifetime, may produce
high energy protons causing thermal runaway and quenching 
of superconducting magnets.\\
In the CERN Large Hadron Collider \cite{LHC} a copper-coated stainless-steel beam pipe (or {\it liner})
is kept at $\approx 20K$ by active Helium cooling, and effectively handles 
the heat load represented by synchrotron radiation, 
photoelectrons, and image-charge losses. 
A large number ($\sim 10^2$  $m^{-1}$) of tiny slots are drilled in the liner wall (see Figure 1)
in order to maintain the desorbed gas densities below a critical level
($\sim 10^{15}$  $molecules/m^3$ for $H_2$) 
by allowing desorbed gas to be continuously cryopumped toward the 
stainless steel cold bore (co-axial to the liner) of the superconducting magnets,
which is kept at $1.9K$ by superfluid Helium.
%
\begin{figure} [!h]
\centerline{ \includegraphics[width=7cm]{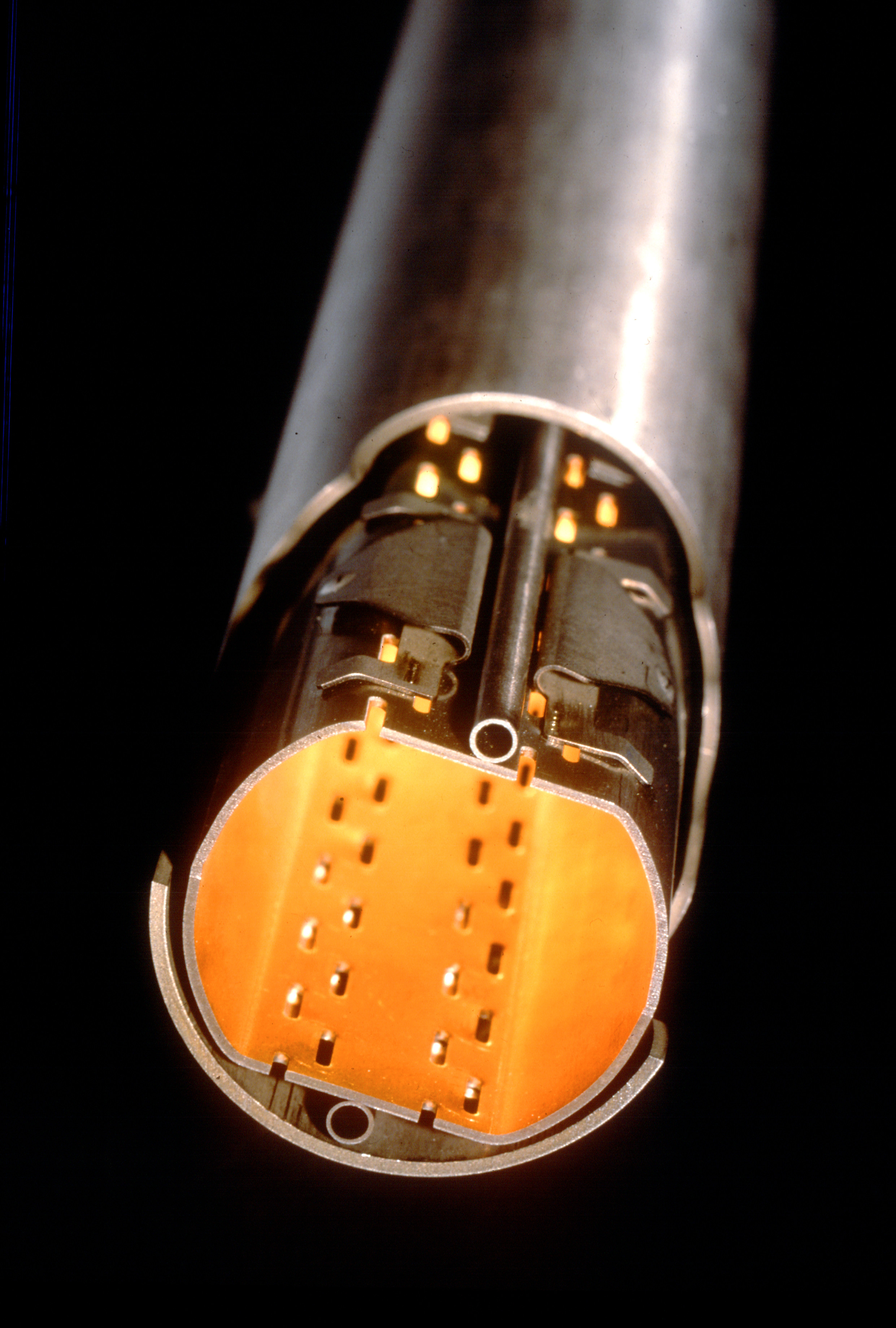}}
\caption{\it
The LHC slotted copper-plated beam pipe and stainless steel cold bore.}
\label{fig1}
\end{figure}
\\
The size, geometry and density of the pumping holes 
affect the beam dynamics and stability  in a way which is
synthetically described by the longitudinal and transverse
beam coupling impedances \cite{Zotter}.  
The hole geometry should be chosen so as to minimize
the effect of trapped (cut-off) modes, 
and the hole pattern should be designed so as to prevent 
the possible coherent buildup of synchrotron radiation 
in the TEM waveguide  limited by the pipe and the cold bore \cite{Heifets}.\\
Open-cell metal foams  could be interesting candidate materials 
for beam liner design. 
In the following we give a brief review of their properties, and of the pertinent modeling tools, 
and draw some preliminary conclusions about the pros and cons of their possible use
in beam liners. 
%
\section{OPEN-CELL METAL FOAMS}
\begin{figure} [!h]
\centerline{ \includegraphics[width=7cm]{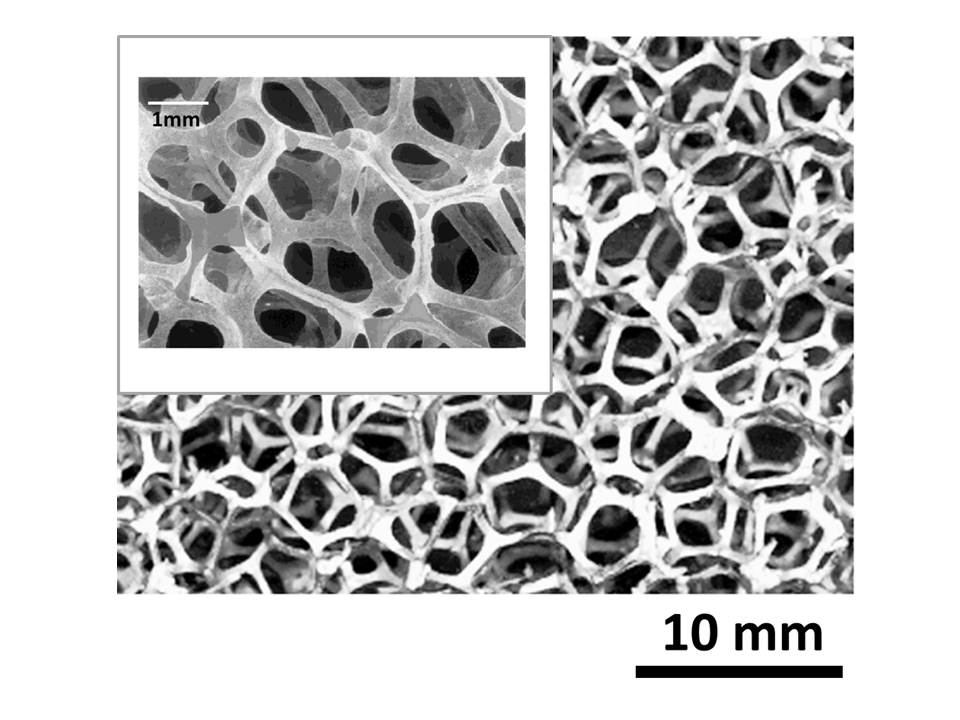}}
\caption{\it
A typical open-cell metal foam structure, at two different viewing scales.}
\label{fig2}
\end{figure}

%
Open-cell metal foams are produced either by vapor- (or electro-) deposition of metal on an open-cell polymer template, 
followed by polymer burn-off, and a final sintering step to densify the ligaments. 
Alternatively, they are obtained by infiltration/casting of molten metal into a solid mould, 
consisting of packed (non-permeable) templates of the pores, 
followed by burn-out and removal of the mould \cite{rev1}. 
Both processes result into highly gas-permeable {\it reticulated} foams, 
where only a 3D web of solid conducting struts among the pores survives. 
The typical structure of these materials is displayed in Figure \ref{fig2}.
%
%
The key structural parameters of reticulated metal foams are the vacuum "pore" size,
and the porosity (volume fraction of pores).
Pore sizes in the range from  $10^{-4}$ to $10^{-3}$ $m$ 
and porosities in the range  0.8 - 0.99  are currently manufactured.
These two parameters determine the material's gas-permeability, 
and, together with the electrical properties of the metal matrix, its electrical
properties.
Metal foams have interesting structural properties 
(low density and weight,  high (tensile and shear)-strength/weight ratio, 
nearly isotropic load response, low coefficient of thermal expansion), 
as summarized in Table I,
which  qualified  them among the most interesting new materials 
for aerospace applications.
%
\begin{figure} [!h]
\centerline{ \includegraphics[width=7cm]{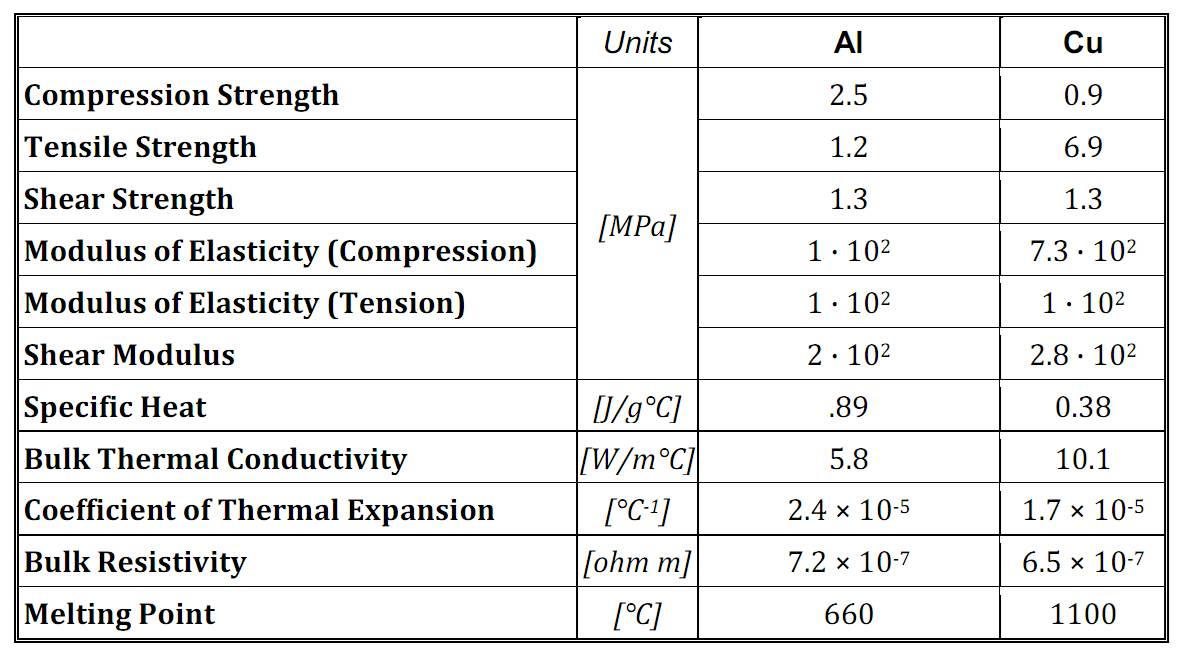}}
\caption{\it
Structural properties of open cell metal foam structure from \cite{ERG}.}
\label{table}
\end{figure}
%
Aluminum and Copper open-cell foams are presently available off-the-shelf \cite{ERG},
and are relatively cheap. 
They can be further coated, e.g., with Silver, Titanium or Platinum, 
for special purpose  applications.
Foams using Steel or Brass, as well as pure Silver, Nickel, Cobalt, Rhodium, Titanium or Beryllium have been
also produced by  a number of Manufacturers. \\
The Weaire-Phelan (WP) space-filling honeycombs are credited 
to provide the  {\it natural}
(i.e., Plateau's  minimal surface principle compliant)
model  of a reticulated metal with equal-sized 
(but possibly unequal-shaped) pores \cite{WeaPh}.
The WP unit cell consists of a certain arrangement of (irregular) 
polyhedra, namely two pentagonal-face dodecahedra (with tetrahedral symmetry $T_h$) ,
and six tetrakaidecahedra (with antiprysmatic symmetry $D_{2d}$) featuring
two hexagonal and twelve pentagonal faces. 
A computer generated WP honeycomb is displayed in Figure \ref{figPW}, and 
its visual similarity to Figure \ref{fig2} is apparent.
%
\begin{figure} [!h]
\centerline{ \includegraphics[width=5cm]{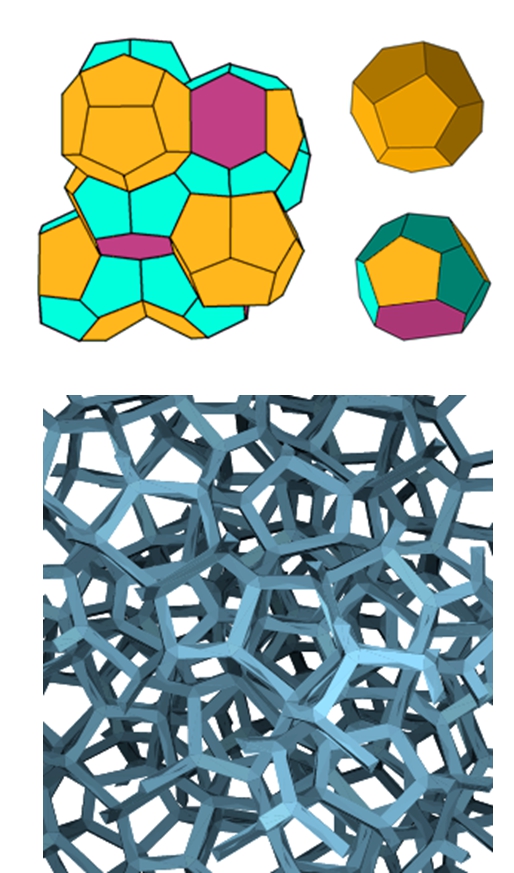}}
\caption{\it
The Wearie-Phelan honeycomb cell (top left), 
its consituent polyhedra (top right),
and  a numerically simulated reticulated foam thereof (bottom).}
\label{figPW}
\end{figure}
%
\subsection{Electrical Properties of Metallic Foams}
\begin{figure} [!h]
\centerline{ \includegraphics[width=7cm]{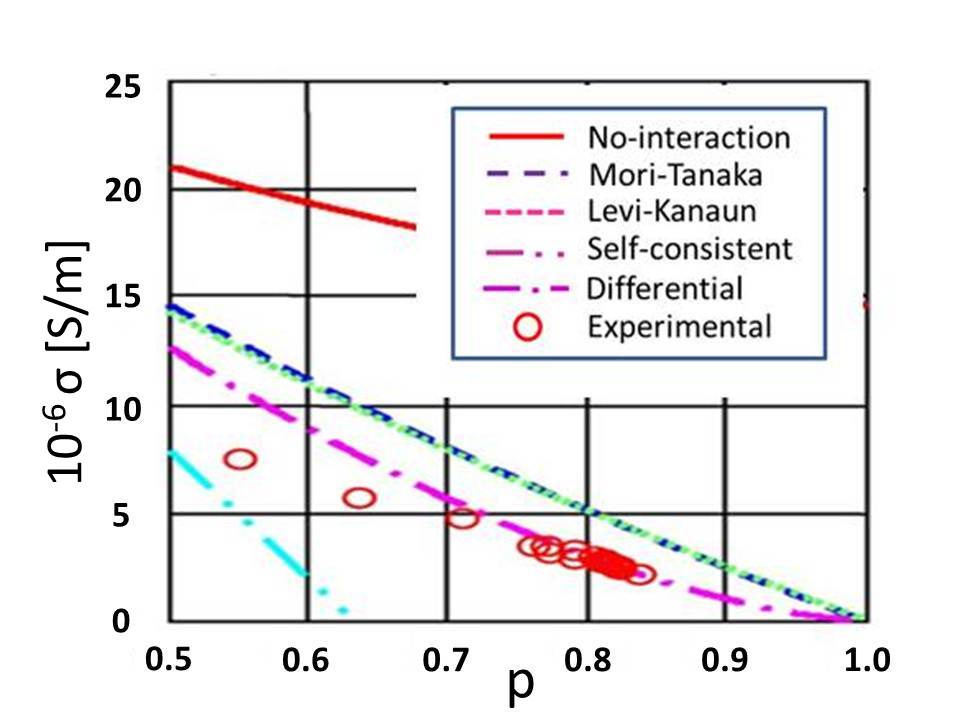}}
\caption{\it
Comparison among different  EMT - based  reticular foam conductivity models.
Static conductivity vs porosity. Aluminum based foam with bulk conductivity
$\sigma_0=3.5\cdot 10^7\mbox{ }S/m$.   
(adapted  from \cite{Sevostianov}).}
\label{fig_comp}
\end{figure}

%
Full electromagnetic modeling of reticulated  metal foams is still to come.
A numerical approach based on Weiland’s finite integration technique
(FIT, \cite{Weiland}) has been proposed by Zhang et al. \cite{Zhang} to compute
the (frequency, thickness and angle of incidence dependent) reflection coefficient 
of  SiC  reticulated foam, and optimize its design.
A simplified model, consisting of stacked square-mesh grids has been used
by Losito et al. \cite{Losito1},\cite{LositoN} to investigate the RF shielding properties
of metallic foams.\\
The main limitation of Zhang's analysis is in the use of a simple body-centered-cubic
unit-cell foam model, for easiest numerical implementation. 
The FIT scheme, however may accommodate in principle more complicated and realistic
foam-cell geometries, including in principle the WP one. \\ 
In the limit where bubbles and metal struts are much smaller than the smallest
wavelength of interest, the DC conductivity of a metal foam can be computed using effective medium theory
(EMT), for which several formulations exist  (see, e.g., \cite{Rev1}-\cite{RevN} for a review).
These include: 
 i) the “infinite dilution” approximation, where inclusions do not interact, 
and are subject to the field which would exist in the homogeneous host;
ii) the self-consistent approach \cite{effective_med}, credited to Bruggemann,
where inclusions are thought of as being embedded in the (yet to be modeled) effective medium;
iii) the differential scheme, whereby inhomogeneities are {\it incrementally} added to the composite\footnote{
In this approach, the total concentration of inhomogeneities does {\it not} coincide with the volume fraction $p$, 
because at each step new inclusions may be placed where old inclusions have already been set.},
until the final concentration is reached, so that at each step the inclusions do not interact, 
and do not modify the field computed at the previous step \cite{differential};
iv) the effective-field methods, whereby interaction among the inclusions is described in terms of an effective field 
acting on each particle, accounting for the presence of the others. 
Two main versions of this method exist, credited to Mori-Tanaka \cite{MorTan} and Levin-Kanaun \cite{LevKan}, 
differing in the way the effective field is computed 
(average over the matrix only, or average over the matrix {\it and} the inclusions, respectively).\\
Both the infinite-dilution and the self-consistent approaches yield 
\begin{equation}
\sigma_{\mbox{\it eff}}=\sigma_0(1-p\nu), 
\label{eq:ID}
\end{equation}
where $\sigma_0$ is the bulk metal conductivity, 
$p$ is the porosity (volume fraction of the vacuum bubbles).
and $\nu$ is a morphology-dependent factor.
The differential approach yields 
\begin{equation}
\sigma_{\mbox{\it eff}}=\sigma_0(1-p)^\nu,
\label{eq:diff}
\end{equation}
while the Mori-Tanaka/Levin-Kanaun approaches yield  
\begin{equation}
\sigma_{\mbox{\it eff}}=\sigma_0/(1+\frac{\nu p}{1-p}).
\label{eq:MT}
\end{equation}
All equations (\ref{eq:ID})-(\ref{eq:MT}) merge, as expected, in the $p \rightarrow 0$ limit.
The various models are synthetically compared in Figure \ref{fig_comp}.
%
%
All these models  predict {\it larger} conductivity then observed in measurements on Al foams\footnote{
It should be noted that open and closed cell metal foams behave similarly in terms of electrical conductivity,
while being markedly different as regards thermal conductivity, due to the different role of convective flow.}. 
This has been attributed to significant oxide formation on the Al conducting web \cite{Clyne}.
Equation (\ref{eq:diff}) agrees in form with predictions 
based on percolation theory \cite{perco} - although strictly speaking
there’s no threshold here beyond which the conducting component disconnects.\\
%
\section{METAL FOAMS vs SOLID METAL\\ PERFORATED WALLS}
\label{sec:vac}
%
In this section we shall attempt to draw a comparison between a metal-foam
beam-pipe wall and a solid-metal perforated one, 
in terms of the relevant vacuum and beam-coupling impedance features .
%
\subsection{Vacuum Issues}
\label{sec:vac}
%
The vacuum dynamics for each molecular species which may be desorbed 
from the wall by synchrotron radiation can be described by
the following set of  (coupled) rate equations \cite{Grobner}
\begin{equation}
\left\{
\begin{array}{l}
\displaystyle{V\frac{dn}{dt}=q-an+b\Theta}\\
\\
\displaystyle{F\frac{d\Theta}{dt}=cn-b\Theta}
\end{array}
\right.
\label{eq:rate}
\end{equation}
Here $n \mbox{ }[m^{-3}]$ and $\Theta \mbox{ }[m^{-2}]$ are the
volume and surface densities of desorbed particles, respectively,
and $V$ and $F$ represent the volume and wall-area of the liner
per unit length, respectively.\\
The first term on the r.h.s. of the first rate equation represents the
number of molecules desorbed by synchrotron radiation per unit length
and time, and is given by
\begin{equation}
q=\eta\dot{\Gamma}
\end{equation}
where $\eta$ is the desorption yield (number of desorbed molecules per incident photon)
and $\dot{\Gamma}$ is the specific photon flux  
(number of photons hitting the wall per unit length and time). 
The second term 
represents the number of molecules which are removed 
per unit time and unit length 
by either sticking to the wall,
or escaping through the holes. 
The $a$ coefficient in (\ref{eq:rate}) can be accordingly written 
\begin{equation}
a=\frac{\langle v \rangle}{4}(s+f)F
\label{eq:aa}
\end{equation}
where $\langle v \rangle\approx(8kT/\pi m)^{1/2}$ is the average molecular speed,
$m$ being the molecular mass, $k$ the Boltzmann constant ant $T$ the absolute temperature, 
$\langle v \rangle/4$ is the average numer of collisions of a single molecule
per unit time and unit wall surface, 
$s$ is the sticking probability,
and $f_h$ is the escape probability. 
%
The third term  
accounts for thermal or radiation induced re-cycling
of molecules sticking at the walls. 
The $b$ coefficient in (\ref{eq:rate}) can be accordingly written
\begin{equation}
b=\kappa\dot{\Gamma}+F \nu_o \exp(-E/kT)
\label{eq:bb}
\end{equation}
Here the first term accounts for radiation induced recycling, 
described by the coefficient $\kappa \mbox{ } [m^{2}]$,
while the second term describes thermally-activated recycling, 
$\nu_0$ being a typical molecular vibrational frequency, 
and $E$ a typical activation energy.\\
The $b\Theta$ term  appears
with reversed sign on the r.h.s. of the second rate equation,
where it represents the the number of molecules {\it de-sticking} from the
wall surface per unit time and unit length.
The first term on the r.h.s. of this equation 
represents the number of molecules 
sticking to the wall, per unit time and unit length, 
whence  (compare with eq. (\ref{eq:aa})) 
\begin{equation}
c=\frac{\langle v \rangle}{4}s F
\label{eq:cc}
\end{equation}
At equilibrium, $\dot{n}=\dot{\Theta}=0$, and the rate equations yield:
\begin{equation}
\left\{
\begin{array}{l}
\displaystyle{n_{eq}=\frac{4\eta\dot{\Gamma}}{\langle v \rangle f F}}\\
\\
\displaystyle{\Theta_{eq}=\frac{s}{f}
\frac{\eta\dot{\Gamma}}{\kappa \dot{\Gamma}+F\nu_o\exp(-E/kT)}
}
\end{array}
\right.
\label{eq:equil}
\end{equation} 
Typical values (from LHC) of the parameters in (\ref{eq:equil})
are collected in Table II below \cite{Grobner}.
%
\begin{figure} [!h]
\centerline{ \includegraphics[width=7cm]{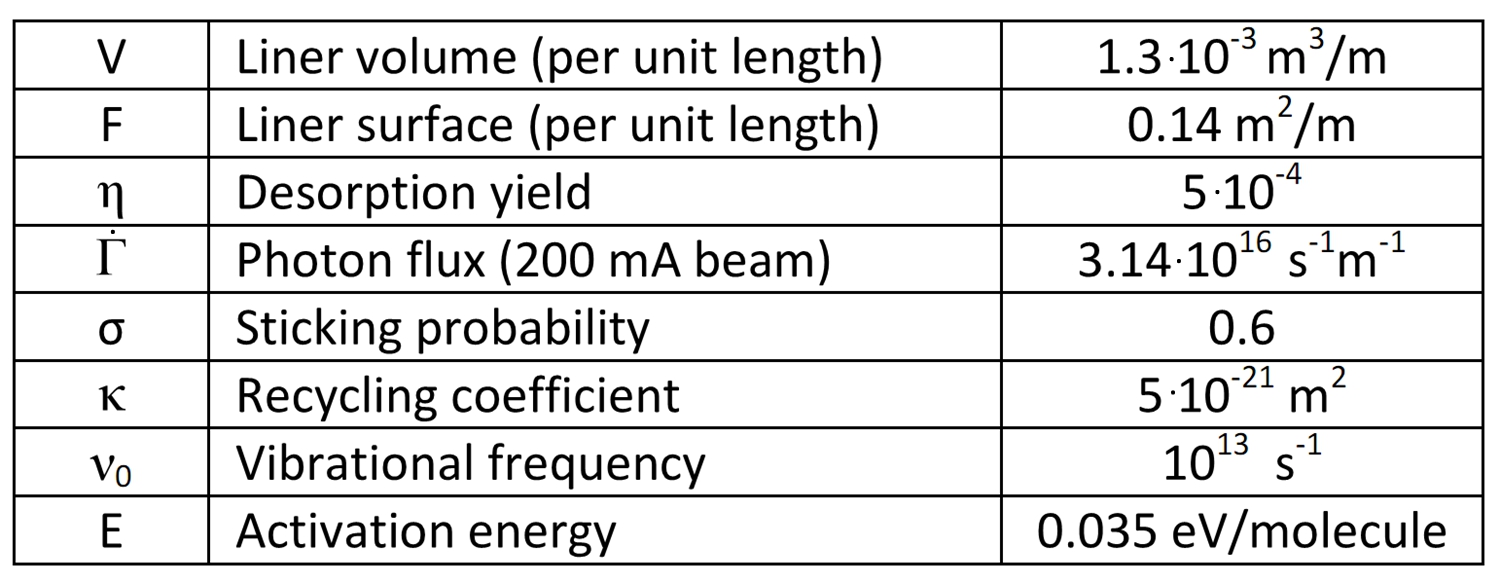}}
\caption{\it
Typical values of the parameters in (\ref{eq:equil}) from \cite{Grobner}.}
\label{table}
\end{figure}
%
The equilibrium molecular densities in (\ref{eq:equil})  should not exceed some {\it critical} values
for safe operation \cite{Grobner}. 
%
\subsubsection{Perforated Solid Metal Wall}
%
For a liner wall with vanishing thickness the escape probability $f$ in
(\ref{eq:aa}) and (\ref{eq:equil}) will be simply equal to the holey fraction $f_h$
of  the wall surface.
For holes drilled in a thick wall, the escape probability
will be {\it less} than $f_h$,  differing from this latter
by a factor $\chi$ (named after Clausing)  which takes into
account the nonzero probability that a molecule may stick at the hole {\it  internal}
surface rather than escaping outside \cite{Steckel}. 
The Clausing factor $\chi$ for thick cylindrical holes 
drilled in a metal wall 
is well approximated by the following empirical formula credited to Iczkowski \cite{Iczko} :
\begin{equation}
 \chi = 1 -0.5(w/R_h)
\label{eq:Claus} 
\end{equation}
where $w$ and $R_h$ are the thickness and radius of the hole.
%
\subsubsection{Open Cell Metal Foam Wall}
%
For an open cell foam the porosity $\rho_h$ (volume fraction of voids),  
average pore radius $R_h$, and volume density of pores $N_h$ 
are related by:
\begin{equation}
\rho_h \approx \frac{4}{3}\pi R^3_h N_h
\end{equation}
which allows to compute $N_h$ from $\rho_h$ and $R_h$. It is reasonable to assume
that the {\it surface} density of the holes will be $\approx N_h^{2/3}$, each hole having an average
surface $\approx (2/3)\pi R_h^2$  so that the fraction of (unit) surface covered by holes will be
\begin{equation}
f_h=0.806\cdot\rho_h^{2/3},
\end{equation}
which will exceed $60\%$ for a typical ($> 0.8$)  (volume) porosities. 
This is a {\it large} number, compared, e.g., to the LHC liner value $f \approx 4.5\%$.\\
This suggests  letting  
$s \rightarrow s(1-f_h)$  
in eqs. (\ref{eq:aa}) and (\ref{eq:equil}), 
since molecules can only stick to the {\it solid} portion of the wall surface. \\
On the other hand, not {\it all} molecules hitting the {\it holey} portion of the wall
will escape, and we may expect  a (much) larger Clausing factor, compared to the simple case
of (right) cylindrical holes drilled in a thick solid plate\footnote{
The gas-permeability of metal foams has been investigated since long, 
both experimentally \cite{Kendall}  and theoretically  (see, e.g., \cite{Boomsma} for a recent account). 
Unfortunately, little attention has been paid so far to the molecular flow limit.} .
We may naively assume that the effective number of molecules (per unit time and length) 
which will {\it escape} from a metal-foam wall with thickness $w$,  
will be related to the number (per unit time and length) of those 
entering the face-holes  by a Lambert-Beers factor, so that
\begin{equation}
f=f_{h}\exp(-w/\ell),
\label{eq:lambert}
\end{equation}
reflecting the fact that those molecules  
may collide with and stick to the (inner) metal web, instead of escaping.  
The obvious requirement that (\ref{eq:lambert}) agrees with (\ref{eq:Claus}) 
in the $w \rightarrow 0$ limit yields  $\ell=2 R_h$  as an estimate 
of the extinction length in (\ref{eq:lambert}).\\
Note that synchrotron radiation will not penetrate the metal foam
beyond a few skin-depths $\delta_S$, so that not {\it all} molecules sticking to the metal web
{\it inside} the metal foam could be recycled by synchrotron radiation. 
This implies that the value of the recycling factor $\kappa$ in (\ref{eq:bb}) and (\ref{eq:equil})
may be significantly different for a reticular wall.
%
\subsection{Beam Coupling Impedance\\ and Parasitic Loss}
\label{sec:conclu}
%
Beam coupling impedances provide a synthetic description
of the beam-pipe interaction, for investigating beam dynamics and stability \cite{Zotter}.
For the simplest case of a circular pipe of radius $b$ with on-axis beam, 
the longitudinal beam-coupling impedance per unit length is given by 
\begin{equation}
   Z_{\parallel}(\omega) = \frac{ Z_{wall} }{ 2\pi b }
\label{eq:Chao}
\end{equation}
where $Z_{wall}$ is the wall impedance, and a Leont\'{o}vich assumption is implied\footnote{
Equation (\ref{eq:Chao}) is a special case of a general formula which allows to compute the longitudinal
and transverse beam-coupling impedances of a pipe with complicated geometric and constitutive properties \cite{Partacc}.}.
%
Similarly, the nonzero components of the diagonal transverse beam coupling impedance dyadic are
\begin{equation}
   Z_{\perp}(\omega) = \frac{ c Z_{wall} }{ \omega \pi b^3 }
\label{eq:ZT}
\end{equation}
where $c$ is the speed of light in vacuum.
The parasitic loss (energy lost by the beam per unit pipe length) is directly
related to the longitudinal impedance via 
\begin{equation}
\Delta{\cal E}=
   \frac{1} {2\pi}
   \int_{-\infty}^{+\infty }|I(\omega)|^2 
   \Re e~Z_{\parallel}(\omega)
   d\omega.
\label{eq:parloss}
\end{equation}
where $I(\omega)$ is the beam-current frequency spectrum \cite{Zotter}.
The beam coupling impedances (and parasitic losses) should not exceed 
some critical values for safe operation \cite{LHC}.\\
%
\subsubsection{Perforated Solid Metal Wall}
%
The wall impedance for a perfectly conducting 
perforated beam pipe was deduced in  \cite{Kurennoy} and \cite{holes} 
in the Bethe limit where the hole size is much smaller than the (shortest) wavelength of interest, yielding 
\begin{equation}
\mbox{Im}[Z_{\parallel}]=-j\frac{Z_0}{2\pi b}
\left(\frac{\omega}{c}\right)
\left( \alpha_e + \alpha_m\right) n_\sigma.
\label{eq:uno}
\end{equation}
\begin{equation}
\mbox{Re}[Z_{\parallel}]=\frac{Z_0}{12\pi^2 b} \left(\frac{\omega}{c}\right)^4
\left( \alpha^2_e + \alpha^2_m\right) n_\sigma
\label{eq:due}
\end{equation} 
where $\alpha_{e,m}$ are the electric and magnetic hole polarizabilities,
and $n_{\sigma}$ is the surface density of holes.
For circular holes with radius $R_h$ 
in a wall with thickness $w$ one has, e.g.,
\begin{equation}
\left\{
\begin{array}{l}
\alpha_e=\frac{2}{3}r^3_0 \exp(-\xi_E w/R_h), \\
\\
\alpha_m=-\frac{4}{3}r^3_0 \exp(-\xi_H w/R_h)
\end{array}
\right.
\end{equation} 
where $\xi_E \approx 2.405$ and $\xi_H \approx 1.841$
are the longitudinal damping constants of the dominant TE and TM
cutoff mode of a circular waveguide having the same radius
$r_0$ as the holes.
Equation (\ref{eq:due}) can be used in (\ref{eq:parloss}) to compute the 
parasitic loss due to the synchrotron radiation leaking through the holes.
The parasitic loss due to the finite bulk conductivity of the liner wall, can
also be obtained from (\ref{eq:parloss}) using 
\begin{equation}
Z_{wall}=\left(
\frac{\omega\mu_0}{\sigma}
\right)^{1/2} \exp(j\pi/4)
\label{eq:Zw}
\end{equation}
in (\ref{eq:Chao}).
%
\subsubsection{Open Cell Metal Foam Wall}
%
The wall impedance of a reticular metal is  given by (\ref{eq:Zw}),
in terms of the {\it effective} conductivity  $\sigma_{\mbox{\it eff }}$ of the material.
Equations (\ref{eq:Chao}) and (\ref{eq:ZT}) give the corresponding beam coupling impedances,  
and equation (\ref{eq:parloss}) yields the related parasitic loss.\\
Heuristically, we can also use the  skin depth of the reticular metal (evaluated at the 
frequency of the synchrotron radiation $\omega_s$) 
\begin{equation}
\delta_S=\left(\frac{2}{\omega_s\mu_0\sigma_{\mbox{eff}}}\right)^{1/2}
\end{equation}
to set the thickness $w$ of the  open-cell metal-foam  wall, 
and estimate  the fraction of parasitic loss 
due to synchrotron radiation leakage as 
\begin{equation}
\Delta{\cal E}^{(rad)}\approx
\Delta{\cal E} \exp(-2w/\delta_S).
\end{equation}
%
\section{CONCLUSIONS}
\label{sec:conclu}
%
On the basis of the above hints, some preliminary qualitative
conclusions can be drawn about the possible use of reticular metals in
beam liners.\\
The structural properties of the material may be  adequate to resist to eddy-current
induced stresses, in case of superconducting magnets'  failure.\\
For a given out-gassing capacity,  a smaller total surface of reticular metal may be needed,
thanks to the much larger gas permeability of open-cell metal foams in the molecular-flow regime,
compared to perforated solid-metal. 
At the same time,  synchrotron radiation leakage could be lower, due to better EM shielding properties, 
and the risk of coherent beaming of synchrotron radiation in the TEM region between the outer liner wall and
the cold bore would be reduced, due to the almost random hole pattern.\\
Bulk ohmic losses in reticular metals, on the other hand, may be much larger compared to solid metals. 
This could be mitigated to some extent by coating the metallic web, e.g., with a superconducting material.\\
Using, e.g., relatively larg(er)  holes/slots in the beam-liner,  
backed by metal foam strips could  possibly  cope with  
the very stringent vacuum and impedance requirements of the perspective SLHC \cite{SLHC}.\\
In order to translate the above hints into quantitative design criteria,  
further modeling effort and substantial experimental work are obviously in order. \\
We believe that such a study program is worth being pursued, 
and that the available modeling tools and technologies provide a good starting point for
its succesful implementation.
We are accordingly preparing a research proposal  on the subject to be 
submitted to  the Italian National Institute
for Nuclear Physics Research (INFN).

\end{document}